\documentclass[11pt]{report}
\usepackage{epsfig}
\usepackage{fancyhdr}
\usepackage{graphics}
\usepackage{graphicx}
\usepackage{dcolumn}
\usepackage{bm}

\usepackage{chbibref}
\setbibref{References}
\usepackage[sort&compress]{natbib}
\usepackage{amsmath}
\usepackage{square_eqn}
\usepackage{booktabs}

\usepackage[normalem]{ulem}
\usepackage{epstopdf}
\usepackage{arydshln}
\usepackage{color}

\usepackage{siunitx}
\usepackage{natbib}
\usepackage{graphicx}
\usepackage{caption}
\usepackage{subcaption}
\usepackage{makecell}
\usepackage{comment}
\usepackage{multirow}
\usepackage{rotating}
\usepackage[labelfont=bf]{caption}

\bibpunct{(}{)}{,}{n}{,}{,}

\begin{document}
\begin{titlepage}
\centering{\textbf{\LARGE{Unconstrained Kinematic MRI Tracking of Wrist Carpal Bones}}}

\vspace{1in}

 \centering{\large{Mohammad Zarenia$^\ast$, Volkan Emre Arpinar, Andrew S. Nencka, L. Tugan Muftuler, and Kevin M. Koch}}

\vspace{0.3in}
 \centering{Department of Radiology, Medical College of Wisconsin, Milwaukee, WI, USA.}\\

\vspace{1in}

\raggedright
$^\ast$Address correspondence to:
\vspace{0.1in}

Mohammad Zarenia, Ph.D.\\
Medical College of Wisconsin\\
Department of Radiology \\
8701 West Watertown Plank Road \\
Milwaukee, WI 53226\\
E-mail: mzarenia@mcw.edu


\end{titlepage}

\section*{Abstract}
In this preliminary study technical methodology for kinematic tracking and profiling of  wrist carpal bones during unconstrained movements is explored. 
Heavily under-sampled and fat-saturated 3D  Cartesian MRI acquisition were used to capture temporal frames of the unconstrained moving wrist of 5 healthy subjects. A slab-to-volume point-cloud based registration was then utilized to register the moving volumes to a high-resolution image volume set collected at a neutral resting position. Comprehensive error analyses for different acquisition parameter settings were performed to evaluate the performance limits of several derived kinematic metrics.
Computational results suggested that sufficient volume coverage for the dynamic acquisitions was reached when collecting 12 slice-encodes at 2.5mm resolution, which yielded a temporal resolution of and 2.57 seconds per volumetric frame. These acquisition parameters resulted in total absolute errors of  1.9$^\circ\pm$1.8$^\circ$ (3$^\circ\pm$4.6$^\circ$) in derived rotation angles and 0.3mm$\pm$0.47mm (0.72mm$\pm$0.8mm) in center-of-mass displacement kinematic profiles within ulnar-radial (flexion-extension) motion.
The results of this study have established the feasibility of kinematic metric tracking of unconstrained wrist motion using 4D MRI.
Temporal metric profiles derived from ulnar-radial deviation motion demonstrated better performance than those derived from flexion/extension movements. Future work will continue to explore the use of these methods in deriving more complex kinematic metrics and their application to subjects with symptomatic carpal dysfunction.  \\\\
\vspace{0.3in}
\noindent {\bf KEY WORDS:} Wrist carpal bones, Registration, 4D MRI, Kinematic 

\setcounter{figure}{0}
\setcounter{table}{0}
\setcounter{equation}{0}

\newpage

\section*{INTRODUCTION}
Imaging of the moving joint has been hypothesized to provide differentiating diagnostic information in orthopedic assessments and longitudinal management \cite{Muhle1999,Johnson2013,Li2004,Teixeira2015,Langner2015,Stromps2018,You2001}. Connective tissue injuries, structural deformities, and structural integrity loss can be difficult to characterize in single-frame static diagnostic images. Improvements in dynamic image acquisition technology and analytic methods continue to offer potential mechanisms to harness the diagnostic information contained in dynamic images of moving joints.  

Dynamic imaging of the wrist is a topic of particular interest in the orthopaedic community. Specifically, it is hypothesized that kinematic analysis of diagnostic images help uncover the role of extrinsic and intrinsic ligaments in wrist dysfunction\cite{SENNWALD1993805,Bateni2013}. 
This concept has been previously explored and utilized to qualitatively identify abnormalities in wrist carpal bone movements  and their correlation with wrist instability \cite{Feipel1999,Teixeira2015,Sikora2019}.
More recent work has also performed preliminary kinematic wrist investigations utilizing x-ray fluoroscopy \cite{Li2004} and CT \cite{Rainbow2013,Best2019,Sikora2019,Roo2019,Zhao2015,Beek2004,Crisco2005}.   

The use of MRI for kinematic analysis of the wrist was first explored by Foster et al, where MRI wrist scans of healthy individuals were acquired at multiple static positions and the carpal bone displacements were studied through a principal component analysis \cite{FOSTER2019}. 
Foster et al's study collected data on an asymptomatic cohort undergoing quasi-static  radial–ulnar deviation movements. Using this data, several metrics, including the scaphotrapezium joint  and the capitate-to-triquetrum distance were derived from the imaging data and analysed to explore modal trends within the study cohort. \cite{Henrichon2020}.

In the present study, we demonstrate an approach utilizing a 4D MRI acquisition of the continuously moving and unconstrained wrist to perform kinematic tracking of individual carpal bones. The acquired dynamic imaging data is registered to a high-resolution complete static volume using a point cloud based registration method.  Metrics derived from these dynamic registrations are then utilized to construct kinematic profiles of the moving wrist.  

In a test cohort of 5 subjects performing unconstrained radial-ulnar and flexion-extension deviations with their dominant hand, kinematic profiles constructed with the described methodology were analyzed for consistency and potential viability in ongoing and future studies. The relative accuracy of the constructed profiles was  examined, and the results for the signal to error analysis of the sample metric profiles were computed.

\section*{METHODS}

The MRI acquisition in this study deploys an unconventional application of the Liver Acquisition with Volume Acceleration (LAVA Flex, GE Healthcare) pulse sequence to capture high resolution static images and 4D kinematic imaging of the moving wrist. 

LAVA Flex is a 3D, Fast Spoiled Gradient Recalled (FSPGR) sequence that collects in-phase and out-of phase echoes for two-point Dixon-based fat-water separation. \cite{LAVAFlex}.  In the present study, the "Water" image output of this technique is utilized for further analysis.   
To our knowledge, this is the first application of the 3D LAVA Flex sequence for the study of joint kinematics using MRI. 
\begin{figure}[]
     \centering
        \includegraphics[width=18cm]{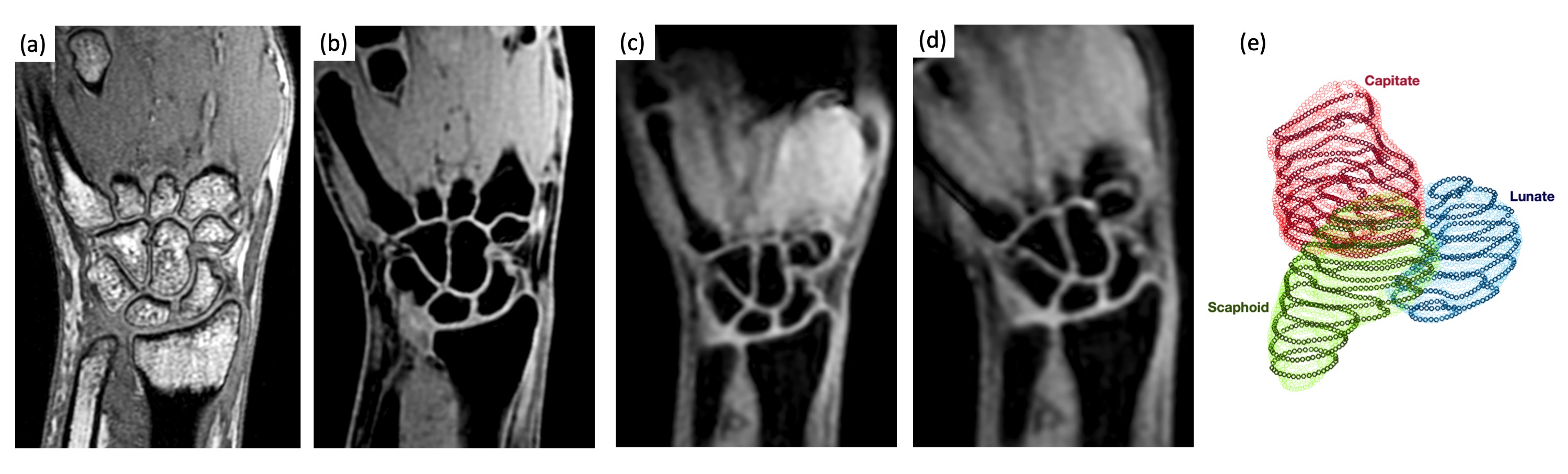}
        \caption{Orthogonal imaging planes through the carpal bones of (a) 3D SPGR and (b) LAVA Flex acquisitions. LAVA Flex  was utilized as a source to manually segment the bones of interest. (c,d) Sample slices of 3D dynamic LAVA Flex images utilized to track the static segmentation. (e) Resulting point cloud of boundary points derived from the high-resolution segmentation. The thick points show the boundaries of the moving slices which are registered to the static point clouds.}
        \label{lava}
\end{figure}
\subsection*{MRI acquisition}
Imaging data were collected on a GE Heathcare Signa Premier 3T MRI scanner using a 16 channel large flex coil. Healthy subjects with no prior known wrist pathology or bone disease were recruited into a locally approved IRB protocol and provided written consent to participate. Subjects were placed in the MRI bore in a prone "superman" position.
The dominant arm of  each subject was placed in the center of the receiver array, using sufficient positioning pads to allow the necessary range of motion for required tasks (i.e. ulnar-radial and flexion-extension deviations in this feasibility study). A picture of the wrist positioned in the scanner is shown in Fig. S1 of the Supplementary Material. No motion-restriction constraints were utilized. Instead, visual cues were used to pace the motion and the subjects were trained prior to the exam using same visual cues (see the training videos in the Supplementary Information).

Static images were acquired using a 3D LAVA Flex sequence with  $0.9\times0.9\times1$mm voxel size and acquisition matrix size of $224\times224\times60$. No acceleration was used for static imaging. Other acquisition parameters were TE $=$ 1.7ms, TR $=$ 5.3ms, FA $=$ 10$^{\circ}$, NEX $=$ 1, and BW$/$Pixel $=$ 417Hz. 
Forty dynamic sub-volumes with a temporal resolution of 2.57s were acquired using multi-phase 3D LAVA Flex series with $1.6\times1.6\times2.5$mm voxel size with $128\times128\times12$ acquisition matrix size. Phase acceleration of 2 and compressed sensing factor of 1.4 were used for dynamic imaging. The other acquisition parameters were TE $=$ 1.2, 3.4ms, TR $=$ 4ms, FA $=$ 10$^{\circ}$, NEX $=$ 1, and BW$/$Pixel $=$ 977Hz.

The visual guidance utilized to direct subject motion indicated 3 cycles of motion during the 103 second acquisition duration. This rate of motion was found to cause minimal motion artifacts in the described dynamic 3D acquisition and was a fast enough rate for subjects to complete without discomfort.

Figures \ref{lava}(a-d) provide demonstrative images from one subject. High resolution static images are shown in \ref{lava}(a) and \ref{lava}(b) obtained using 3D SPGR and LAVA Flex MRI acquisitions, respectively. The LAVA Flex water (i.e. fat-suppressed) reconstructed image series provided excellent bone/tissue contrast, which is essential for accurately segmenting the carpal bones. Two sample images from the 3D dynamic LAVA Flex series during an ulnar-radial deviation are shown in Figs. \ref{lava}(c) and \ref{lava}(d). 
\subsection*{Registration approach}
The average number of slices acquired during the dynamic imaging sequence of ulnar-radial (flexion-extension) motion covered respectively 88$\pm$7.8 (70.1$\pm$9.8), 95$\pm$7.7 (98$\pm$12.9), and 90$\pm$7.2 (83$\pm$9.7) percent of the scaphoid, capitate, and lunate. These values are obtained by averaging over 200 time frames of 5 subjects, i.e. 40 acquired frames for each subject. 

The mapping of sub-volume images to a given volumetric (3D) reference image is a challenging image registration problem \cite{FERRANTE2017}. 
Here, this problem is addressed using a point-cloud based registration algorithm on manually segmented carpal bones. The manual segmentation was done on both the high resolution static images and the dynamic image volumes using the OHIF viewer \cite{OHIF}.

Registration of dynamic to static point clouds derived from the OHIF RTStruct DICOM segmentation outputs was performed in MATLAB \cite{matlab} using its iterative-closest-point registration function, {\it pcregistericp}.  This function accepts the moving and static point clouds as inputs and performs a 6-degrees of freedom rigid-body transformation by minimizing the root-square-mean-square error (RMSE) of the Euclidean distance between the coordinates of the matched pairs defined by, 
\begin{equation}
{\rm RMSE}=\left(\frac{\sum_{i=1}^{N_m}|{\bf R}_m-{\bf R}_f|^2}{N_m}\right)^{1/2},
\label{rmse}
\end{equation}
where, ${\bf R}_m$ and ${\bf R}_f$ are the coordinates of the nearest pairs in the 
moving and static point clouds, respectively. $N_m$ is number of boundary points of the moving point cloud. 

The outputs of the registration function are a $4\times4$ transformation matrix, the registered point cloud, and the resulting RMSE. 
The transformation matrix consists of a $3\times3$ rotation sub-matrix and 3 spatial transformation components where the rotation elements are represented in an angle-axis form. 
Figure \ref{lava}e provides an example surface point cloud  of the moving slab (thick points) that has been registered to the surface point cloud of the static volume.
\begin{figure}[]
     \centering
        \includegraphics[width=17cm]{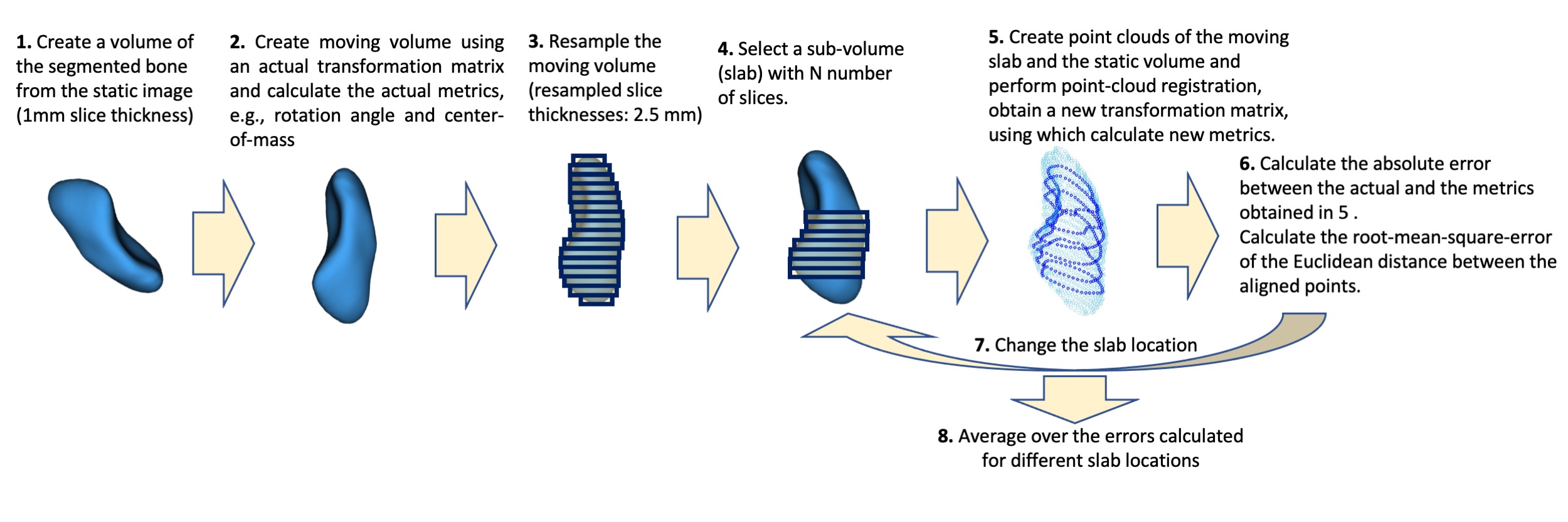}
        \caption{Workflow for processing of error analysis in each time frame.}
        \label{flowchart}
\end{figure}
\section*{Error analysis}
The error analysis process used to calculate the registration error and the resulting absolute errors in reported metrics is depicted in a flowchart in Figure \ref{flowchart}.
\begin{figure}[]
     \centering
        \includegraphics[width=14cm]{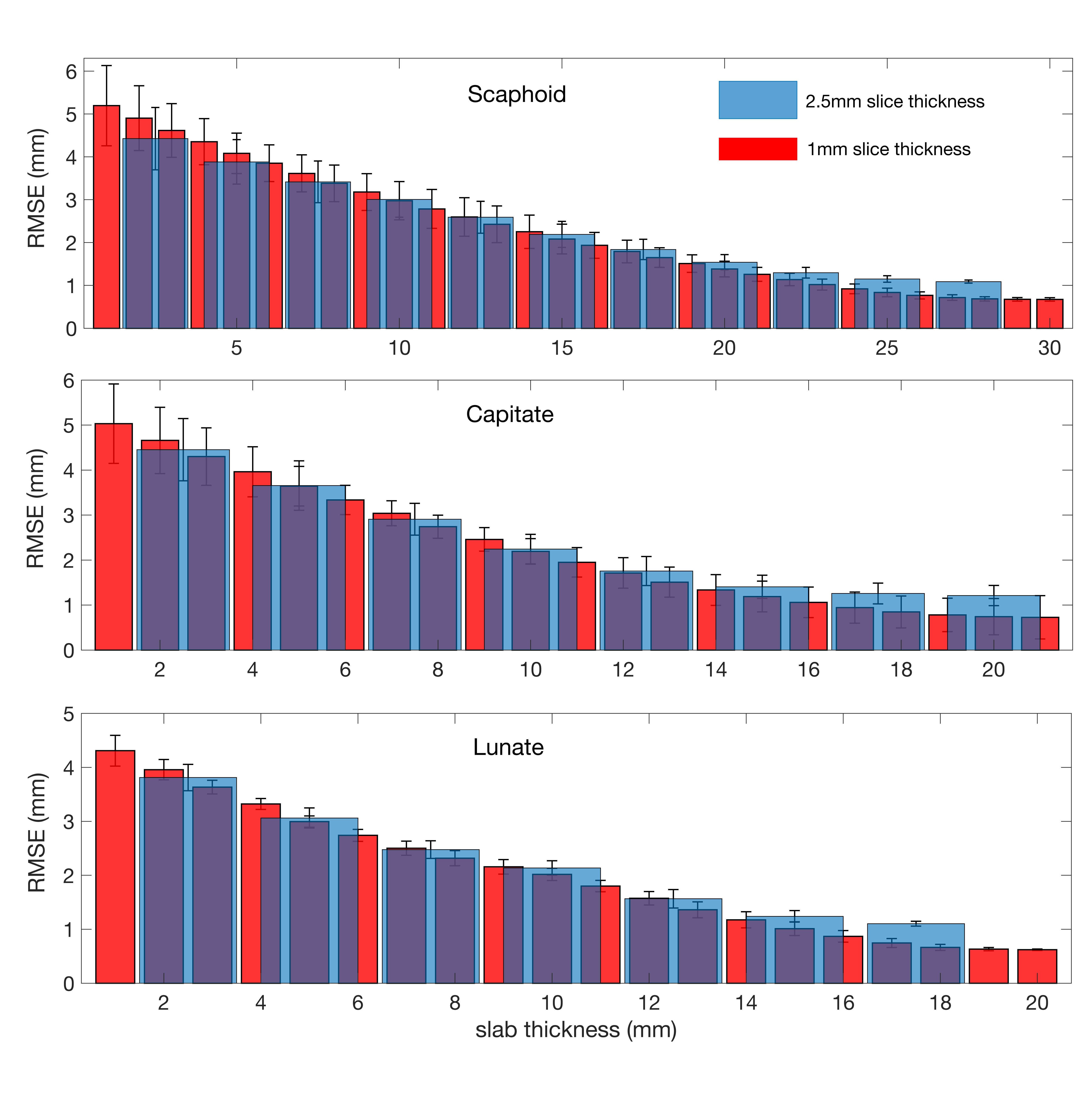}
        \caption{Root-Mean-Square-Error (RMSE), Eq. (\ref{rmse}), averaged over 10 different time frames and for all possible slab locations as functions of slab thickness  for 1mm and 2.5mm slice thicknesses of moving slabs of scaphoid, capitate, and lunate. Error bars indicate the standard deviation.}
        \label{error}
\end{figure}
The high-resolution  static volume is first transferred to the moving reference frame using the inverse of the transfer matrix obtained through the registration of the segmented moving slab to the static volume. The resulting moving volume is then resampled to create a moving volume with 2.5mm slice thicknesses (i.e. the same slice thickness as the actual moving slabs).
From the resampled moving volume, a sub-volume with $N$ slices is chosen ($N$ is the number of segmented slices in actual moving images). The value of $N$ and location of sub-volume with respect to the full bone was varied for this analysis. Using this synthetically generated sub-volume, a new registration matrix was calculated with respect to the original high-resolution image. The absolute errors were calculated between the metrics derived from the actual motion and this synthetic data, such as rotation angles and center-of-mass, as well as the RMSE between the registered and moving slab.
\begin{figure}[]
\vspace{-1.5cm}
     \centering
        \includegraphics[width=18cm]{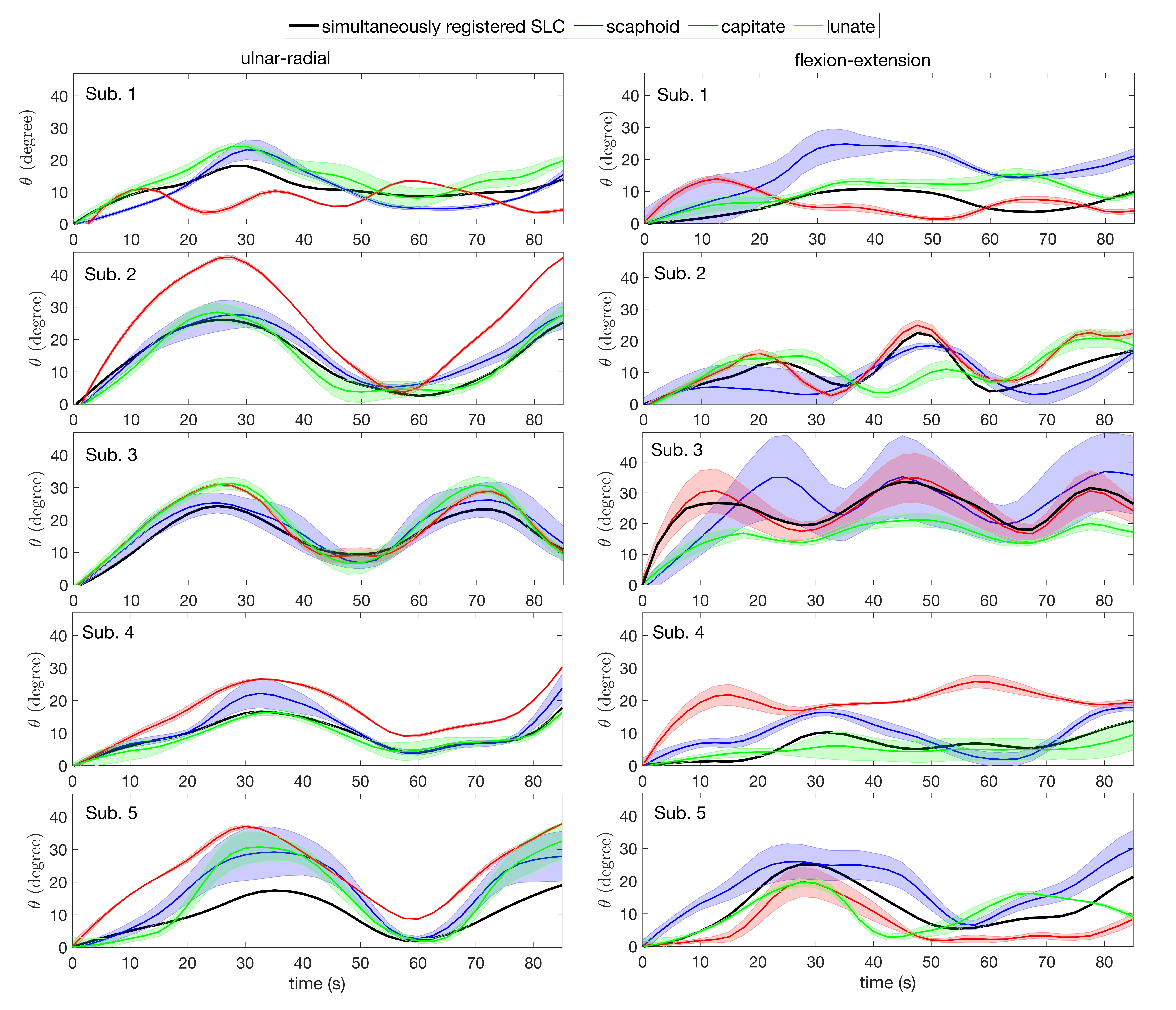}
        \caption{Kinematic profiles of scaphoid, capitate, and lunate rotation angles in relation to their starting positions. The black solid curve represents the profile of simultaneously registered bones. The absolute errors are represented by the shaded error bands, calculated using the process shown in Fig. \ref{flowchart}}
        \label{angles}
\end{figure}

\begin{figure}[]
     \centering
    \includegraphics[width=12cm]{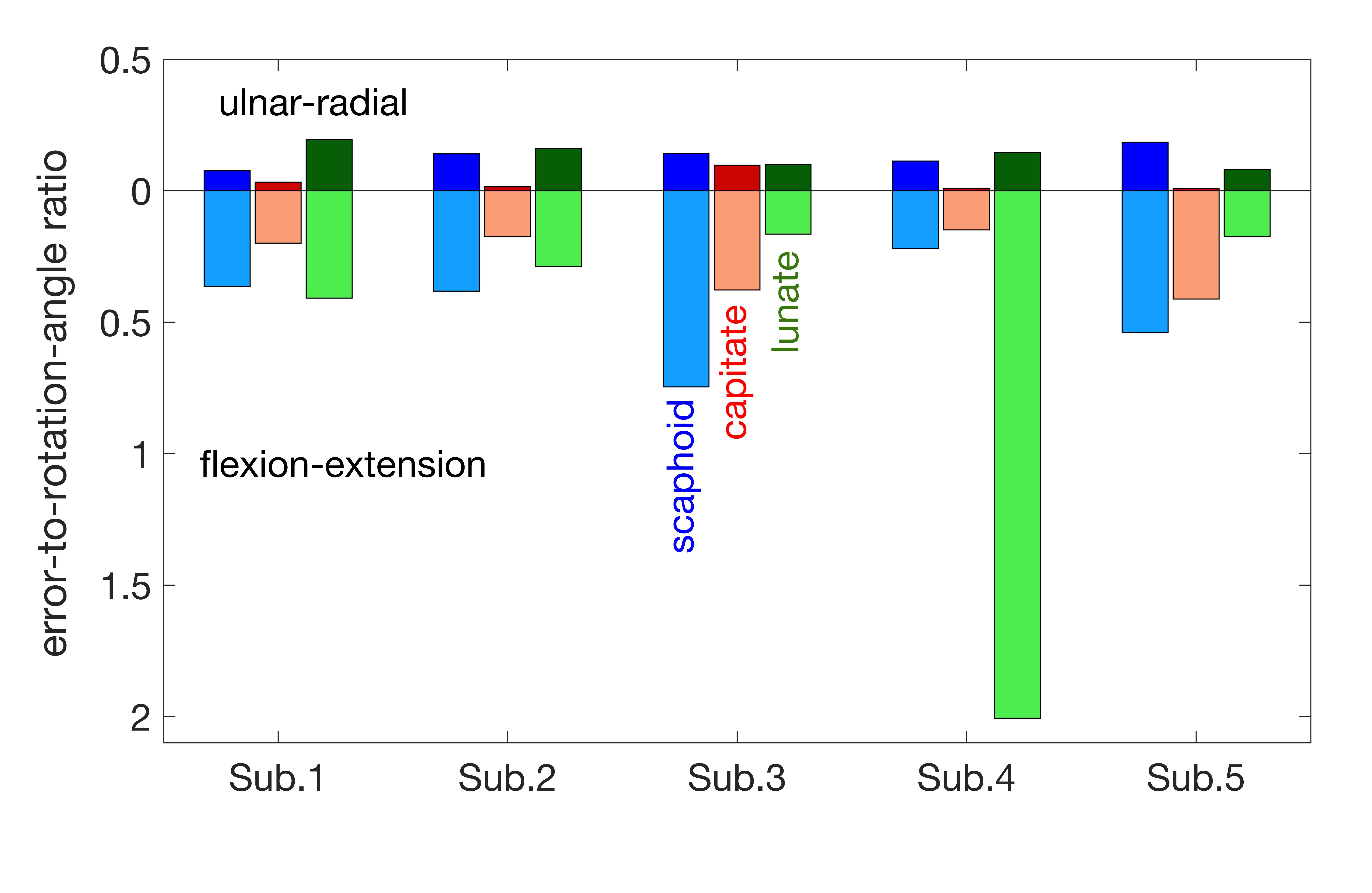}
        \caption{Error-to-signal ratio of rotation angle profiles (absolute-error$/\theta$) that are obtained by averaging over the ratio values at the extremum points of the profiles of Fig. \ref{angles}.}
        \label{errorToangle}
\end{figure}

\section*{RESULTS}
To characterize the performance of the registration as a function of the slice thickness and number of collected slices (i.e. the collected slab thickness), Fig. \ref{error} compares the averaged RMSEs over 10 different kinematic position frames for all possible sub-volume slab locations within the bone at each frame.

As the slab thickness is increased, the registration performance improves. These findings also demonstrate that slab thickness (bone coverage) has a greater impact on registration performance than slice resolution. The RMSE difference between slabs with 1mm and 2.5 slice thicknesses is found to be less than 0.5mm (i.e. less than the 1mm slice thickness of the static volume) at slab thicknesses with more than 70$\%$ bone coverage  -- which is the case for the data collected and analyzed in this pilot study. 

Figure \ref{angles} provides derived  kinematic metric profiles for the carpal bone rotation angles with respect to the initial hand position. To obtain the rotation angle with respect to the initial position, the high-resolution static volume was first registered to the initial moving slab and the resulting volume was then utilized to register the other moving slabs. The shaded bands are the absolute errors evaluated through the process demonstrated in Fig. \ref{error}. During the error analysis, $N$ was fixed to the number of segmented slices and all possible slab locations with respect to the full volume were considered at each time frame. In order to simplify the profile and clarify the modal nature of the motion, a smoothing filter is applied. The smoothing was performed in MATLAB using {\it rloess} method with a span value of 20\% of the total number of points. This method assigns zero weight to data outside six mean absolute deviations. To compare the carpal bone's rotation with the wrist, rotation angles of simultaneously (i.e. bulk) registered scaphoid, capitate, and lunate (black curves) are superimposed. The maximum rotation angles for the scaphoid, capitate, and lunate in ulnar-radial (flexion-extension) motion were found to be $28.6^\circ\pm 3.4^\circ$ ($27^\circ\pm 7.8^\circ$), $34.6^\circ\pm 13.2^\circ$ ($24.6^\circ\pm 7.2^\circ$), and $31.4^\circ\pm 12^\circ$ ($17.5^\circ\pm 6.8^\circ$), represented as  means and standard deviations of 5 subjects. 
An error-to-signal ratio for the rotation angle profiles, i.e., (absolute-error)$/\theta$, is provided in Fig. \ref{errorToangle}. For the ulnar-radial (flexion-extension) motion, the average ratios of 5 subjects for scaphoid, capitate, and lunate are respectively 0.26$\pm$0.08 (0.57$\pm$0.21), 0.06$\pm$0.07 (0.27$\pm$0.12), 0.29$\pm$0.06 (0.57$\pm$0.66).
\begin{figure}[]
     \centering
        \includegraphics[width=16cm]{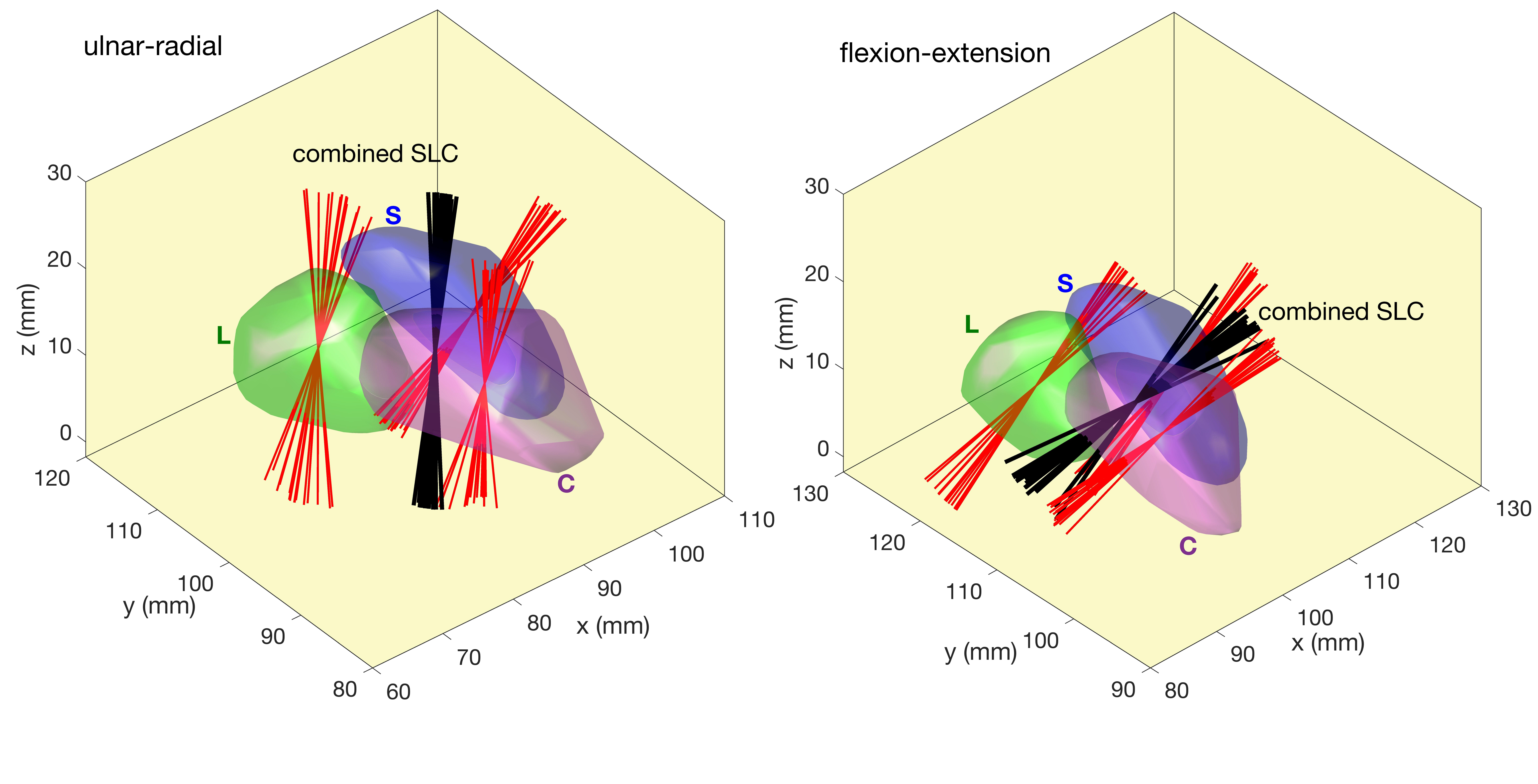}
        \caption{Scaphoid, capitate, lunate (red) and their simultaneously registered (black) angular axes of rotation of 15 time frames corresponding to 10s$<$time$<$62s for subject 3.}
        \label{axes}
\end{figure}

Figure \ref{axes} provides exemplary angular axes of rotation over 15 time frames (corresponding to 10s$<$time$<$62s) of scaphoid, capitate, and lunate  as well as the results of their simultaneous registration in subject 3. In each case, the rotation axes pass through the center-of-mass. 
As expected, the principal axes of rotation of simultaneously-registered bones (black lines in Fig. \ref{axes}) are directed along the rotation axes of the wrist motion that is perpendicular to the orthogonal plane during an ulnar-radial deviation and parallel to the axial plane during flexion-extension. 
\begin{figure}[]
     \centering
     \vspace{-2cm}
     \includegraphics[width=8cm]{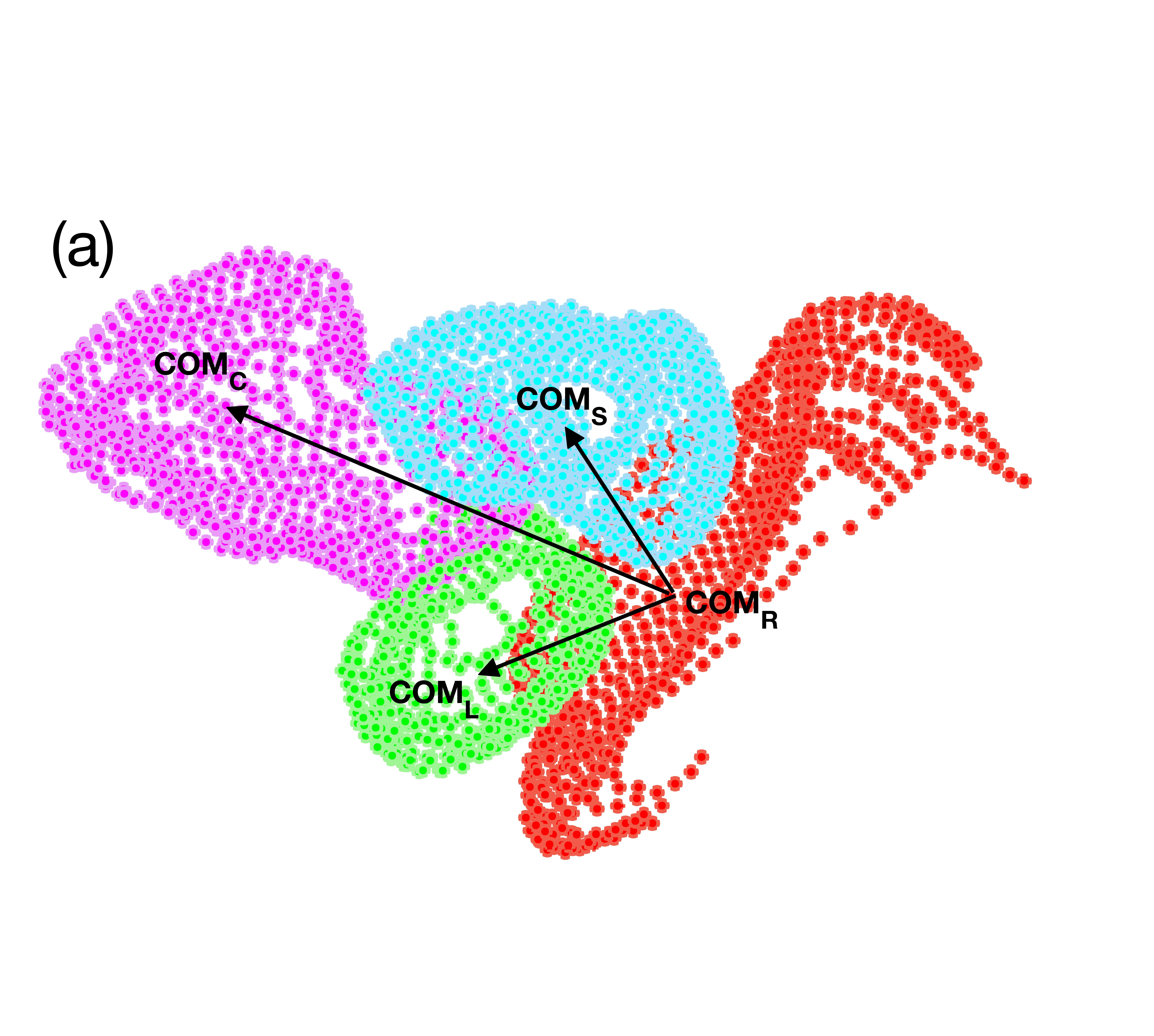}\\
     \vspace{-1cm}
     \centering
    \includegraphics[width=17cm]{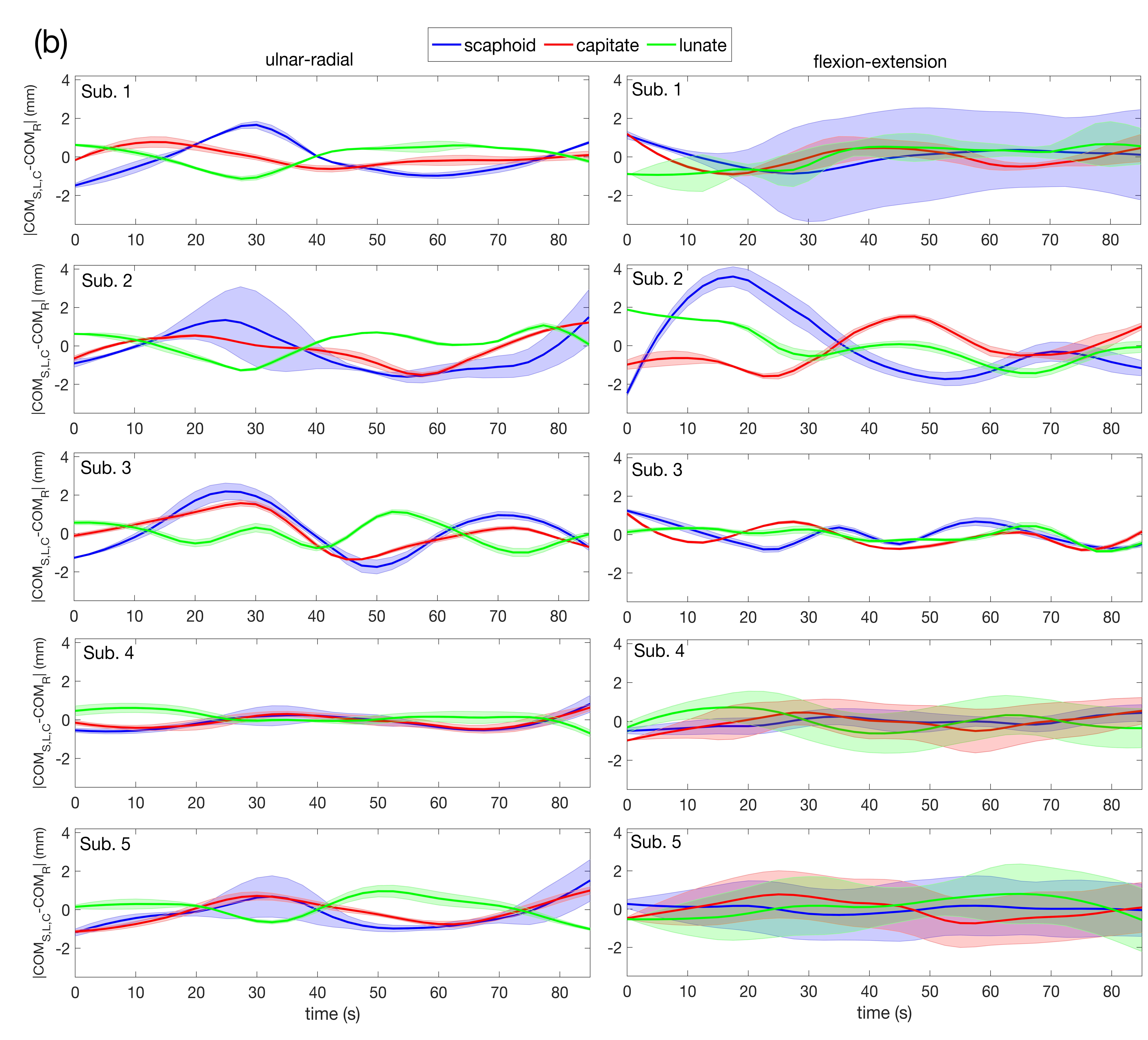}
        \caption{(a) Point clouds of scaphoid, capitate, lunate, and articular surface of radius. (b) Kinematic profiles of the scaphoid, capitate, and lunate center-of-masses with respect to the radius center-of-mass. 
        The mean value is subtracted from each profile.
        The shaded error bands are the absolute errors, calculated using the process shown in Fig. \ref{flowchart}}.
        \label{com}
\end{figure}

\begin{figure}[]
     \centering
    \includegraphics[width=12cm]{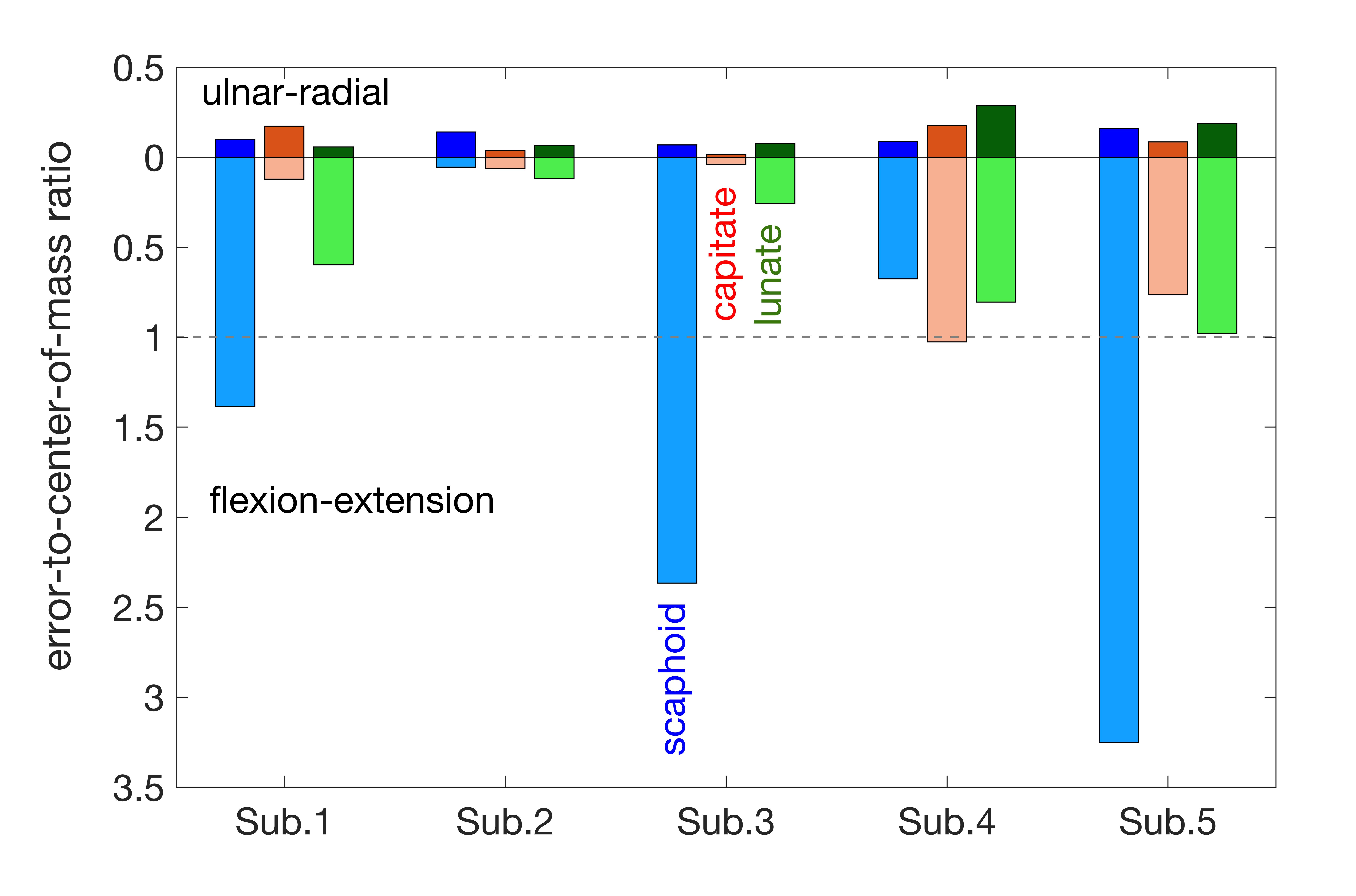}
        \caption{Error-to-signal ratio of center-of-mass profiles (absolute-error$/$center-of-mass) that are obtained by averaging over the ratio  values at the extremum points of the profiles of Fig. \ref{com}.}
        \label{errorToCOM}
\end{figure}

Kinematic profiles of the center-of-mass displacement of carpal bones with respect to radius are shown in Fig. \ref{com}.
To derive these profiles, the articular surface of the radius was segmented out of the static and moving images and registered using the discussed point cloud approach. 
The center-of-mass profiles of simultaneously registered scaphoid, capitate, and lunate are compared to the combined canter-of-mass of individually registered bones in Fig. S3 of the Supplementary materials. As expected, the results are close to one other, indicating logically consistent registration performance.
Averaging over 200 time frames of 5 subjects, the range of variations (mean $\pm$ standard deviation) in carpal bones-radius center-of-mass distances during the ulnar-radial (flexion-extension) motion are respectively obtained as 16.29$\pm$1.05 mm (15.98$\pm$0.63 mm), 24.44$\pm$0.63 mm (23.55$\pm$0.6 mm), and 11.53$\pm$0.62 mm (11.71$\pm$0.58 mm)) for scaphoid, capitate, and lunate. 
Fig. \ref{errorToCOM} provides the results for the error-to-signal ratio of center-of-mass profiles (absolute-error$/$center-of-mass) which are calculated at the extremum points. The average ratios of 5 subjects for the ulnar-radial (flexion-extension) motion are 0.11$\pm$0.03 (1.54$\pm$1.28), 0.1$\pm$0.075 (0.4$\pm$0.45), 0.13$\pm$0.09 (0.55$\pm$0.36) for the scaphoid, capitate, and lunate, respectively.

Table \ref{tableError} provides error analysis results of the rotation angle and center-of-mass metrics as functions of temporal coverage and number of acquired moving slices in each time frame. 
Using the fewer number of acquired slices at temporal resolutions greater than 2.57s/frame, registration errors were significantly increased.  For poorer temporal resolutions using higher numbers of slices per volume, the registration accuracy slightly improved (as expected), however empirical testing identified difficulties for subjects to move their wrists at such a slow rate.  Therefore, the utilized approach
(12 2.5mm slices acquired at 2.57s/frame ) provided an optimal balance of subject comfort, image/data quality, and registration accuracy.      
%

\begin{sidewaystable}
\caption{Absolute errors of rotation angle and center-of-mass metrics for different acquired slices and temporal resolutions. Due to slice overlapping (to prevent foldover aliasing artifacts), 4 of the initially acquired slices in each volume were removed. The preliminary kinematic profiling of carpal bones in this study was performed using 12 slices (out of 16 initially acquired slices).  This case is highlighted in bold text.   The error in each case is obtained by averaging errors computed over the 200 time frames of 5 subjects.}
\vspace{0.2in}
\centering
\small
\begin{tabular}{|l|l|lll|lll|lll|lll|}
\hline
\hline
\multirow{3}{*}{\begin{tabular}[c]{@{}l@{}}acquired\\ slices\end{tabular}} & \multirow{3}{*}{\begin{tabular}[c]{@{}l@{}} temporal\\ resolution \\~~~(s)\end{tabular}} & \multicolumn{6}{l|}{~~~~~~~~~~~~~~~~~~~~~~~rotation angle error (degree)} & \multicolumn{6}{l|}{~~~~~~~~~~~~~~~~~~~~~~~~~center-of-mass error (mm)} \\ \cline{3-14} 
 &  & \multicolumn{3}{l|}{~~~~~~~~~~~ulnar-radial} & \multicolumn{3}{l|}{~~~~~~~~~flexion-extension} & \multicolumn{3}{l|}{~~~~~~~~~~~~ulnar-radial} & \multicolumn{3}{l|}{~~~~~~~~~flexion-extension} \\ \cline{3-14} 
 &  & ~~S & ~~C & ~~L & ~~S & ~~C & ~~L & ~~S & ~~C & ~~L & ~~S & ~~C & ~~L \\ \hline
~~~~~4 & ~~~~~~2.1
  & 30.5$\pm15.6$ & 17.7$\pm12.3$ & 50$\pm15.6$ & 42.8$\pm$19.2 & 19.5$\pm$16.5 & 56.5$\pm$19.2 & 2.28$\pm$1.3 & 2.6$\pm$2.3 & 4.01$\pm$2.6 & 3.4$\pm$2.2 & 2.7$\pm$2.1 & 3.8$\pm$2.3 \\ 
~~~~~8 & ~~~~~~2.3  & 6.3$\pm3.8$  & 1.8$\pm1$  & 10.1$\pm8$ & 21.6$\pm$14.9 & 4.05$\pm$2.4  & 16.06$\pm$15.4 & 2.4$\pm$2.1 & 0.38$\pm$0.42 & 1.11$\pm$0.9 & 1.6$\pm$1.1 & 1.07$\pm$0.7  &1.02$\pm$1.1  \\ 
~~~~~{\bf 12} & ~~~~~~{\bf 2.57} & {\bf 3.6$\pm$2.7}  & {\bf 0.61$\pm$0.38}  & {\bf 2.7$\pm$1.9}  & {\bf 7.17$\pm$ 5.2}  & {\bf 2.2$\pm$+1.9}  & {\bf 2.4$\pm$1.7}  & {\bf 1.04$\pm$0.85} & {\bf 0.3$\pm$0.18} & {\bf 0.97$\pm$0.8} & {\bf 1.41$\pm$1.4} & {\bf 0.62$\pm$0.47} & {\bf 0.67$\pm$0.36} \\ 
~~~~~16 & ~~~~~~3.27 & 2.35$\pm1.9$  & 0.54$\pm0.38$  & 2.12$\pm1$  & 3.1$\pm$2.6  & 2.05$\pm$1.9  & 1.7$\pm$0.98  & 0.77$\pm$0.5 & 0.23$\pm$0.17 & 0.73$\pm$0.6 & 1.0$\pm$0.64 & 0.60$\pm$0.48 & 0.42$\pm$0.34  \\ 
~~~~~20 & ~~~~~~3.72 & 2.16$\pm1.9$  & 0.58$\pm0.38$  & 2.1$\pm1$  & 2.7$\pm$2.2  & 2.05$\pm$1.9  & 1.7$\pm$0.97  & 0.77$\pm$0.5 & 0.23$\pm$0.17 & 0.7$\pm$0.6 & 0.97$\pm$0.65 & 0.60$\pm$0.48 & 0.40$\pm$0.28 \\ \hline\hline
\end{tabular}
\label{tableError}
\end{sidewaystable}
%
\normalsize	
\section*{DISCUSSION}
In the present study, the motions of scaphoid, capitate, and lunate carpal bones within 5 subjects were analyzed during unconstrained ulnar-radial deviation and flexion-extension movements. Nine basic metrics were analyzed as  proof-of-concept kinematic metric profiles: 6 angular rotation angles and axes and 3  center-of-mass displacements of carpal bones with respect to the radius.  

Individual carpal bone rotation angles followed similar profiles as the bulk composite of the 3 bones within the ulnar-radial motion (Fig. \ref{angles}).
An interesting exception was observed in subject 1, whereby  the capitate displayed a different profile compared to the other subjects and a significantly decreased range of rotation ($0<\theta <13^{\circ}$) was seen. The registration performance in this bone/subject was inspected and found to be accurate. After an examination of other high resolution static images of this subject's wrist, a structural abnormality was detected in capitate (see Fig. S2 in the Supplementary Material). 
Although further examination is needed to determine the  nature and impact of this abnormality, this unexpected finding is noteworthy. 

The center of mass kinematic profiles  (Fig. \ref{com}) suggest that the scaphoid and capitate move mainly in-phase, while the lunate moves out of phase (relative to the proximal radius surface) during the ulnar-radial motion.  Again, there was an exception to this trend for the capitate profile of subject 1. 

The kinematic profiles of flexion-extension motion, showed that range of rotation of the carpal bones were lower in capitate and lunate by about 30\% and 45\%, respectively, compared to their range of rotation within ulnar-radial motion. However, the range of center-of-mass displacement relative to radius remains comparable (less than 5\% difference) during the both ulnar-radial and flexion-extension motions.

The results of this feasibility study are encouraging.  First, the study has demonstrated an ability to extract repeatable kinematic metric profiles of carpal bones from 4D MRI of unconstrained wrist motion.  Of the 9 studied metrics, the individual rotation angles of carpal bones within the ulnar-radial deviation showed the most promise as  normative MRI-derived kinematic wrist profiles. The kinematic metrics derived from  flexion-extension motions demonstrated more variation and lower signal to error signatures in this preliminary study.  However, more complex metric derivations may improve this performance of the flexion-extension analysis.  

Quantitative analysis of unconstrained wrist motion using MRI offers a rich set of potential tools that could be utilized to study joint degeneration and dysfunction. The relatively fast kinematic profiling MR acquisitions demonstrated in this feasibility study could  be added to routine orthopedic MRI exams with relative ease, therefore providing optimal diagnostic imaging in both static and kinematic contexts in a single visit.
Of note, the unique acquisition approach demonstrated in this study does not use any prototype pulse-sequences and is commercially available on most GE Healthcare scanner platforms.   Similar sequences and capabilities are also available on other vendor platforms.  As a result, the acquisition methods demonstrated in this work could easily be implemented within many clinical imaging environments.    

In addition to profiling of rotation angles and center-of-mass displacements of scaphoid, lunate, capitate and their combined metrics,  many more metrics such as gaps between the carpal bones, rotation angles of carpal bones relative to radius, etc., across several more wrist motions can be tracked in similar fashion. Future work will continue to deploy these new technologies to explore more complete kinematic tracking and profiling of unconstrained wrist motion.
\section*{ACKNOWLEDGEMENTS}
Research reported in this publication was supported in part by National Institute of Health (NIH) grant R21AR075327. 
\section*{ETHICAL APPROVAL}
All procedures performed in studies involving human participants were in accordance with the ethical standards of the institutional and/or national research committee and with the 1964 Helsinki declaration and its later amendments or comparable ethical standards.\\
{\it {\bf Informed Consent:}} Informed consent was obtained from all individual participants included in the study.\\
\section*{Conflict of Interest}
The authors declare that they have no conflicts of interest.
\newpage
\section*{FIGURE LEGENDS}
\noindent \textbf{Figure 1:} Orthogonal imaging planes through the carpal bones of (a) 3D SPGR and (b) LAVA Flex acquisitions. LAVA Flex  was utilized as a source to manually segment the bones of interest. (c,d) Sample slices of 3D dynamic LAVA Flex images utilized to track the static segmentation. (e) Resulting point cloud of boundary points derived from the high-resolution segmentation. The thick points show the boundaries of the moving slices which are registered to the static point clouds.

\vspace{0.1in}
\noindent \textbf{Figure 2:} Workflow for processing of error analysis in each time frame.

\vspace{0.1in}
\noindent \textbf{Figure 3:} Root-Mean-Square-Error (RMSE), Eq. (\ref{rmse}), averaged over 10 different time frames and for all possible slab locations as functions of slab thickness  for 1mm and 2.5mm slice thicknesses of moving slabs of scaphoid, capitate, and lunate. Error bars indicate the standard deviation.

\vspace{0.1in}
\noindent \textbf{Figure 4:} Kinematic profiles of scaphoid, capitate, and lunate rotation angles in relation to their starting positions. The black solid curve represents the profile of simultaneously registered bones. The absolute errors are represented by the shaded error bands, calculated using the process shown in Fig. \ref{flowchart}

\vspace{0.1in}
\noindent \textbf{Figure 5:} Error-to-signal ratio of rotation angle profiles (absolute-error$/\theta$) that are obtained by averaging over the ratio values at the extremum points of the profiles of Fig. \ref{angles}.

\vspace{0.1in}
\noindent \textbf{Figure 6:} Scaphoid, capitate, lunate (red) and their simultaneously registered (black) angular axes of rotation of 15 time frames corresponding to 10s$<$time$<$62s for subject 3.

\vspace{0.1in}
\noindent \textbf{Figure 7:} (a) Point clouds of scaphoid, capitate, lunate, and articular surface of radius. (b) Kinematic profiles of the scaphoid, capitate, and lunate center-of-masses with respect to the radius center-of-mass. The mean value is subtracted from each profile. The shaded error bands are the absolute errors, calculated using the process shown in Fig. \ref{flowchart}.

\vspace{0.1in}
\noindent \textbf{Figure 8:} Error-to-signal ratio of center-of-mass profiles (absolute-error$/$center-of-mass) that are obtained by averaging over the ratio  values at the extremum points of the profiles of Fig. \ref{com}.

\newpage
\section*{SUPPLEMENTAL MATERIAL}
%
\renewcommand{\thefigure}{S\arabic{figure}}
\setcounter{figure}{0}
\setcounter{table}{0}
\setcounter{equation}{0}
%

\noindent {\bf Videos of ulnar-radial (Video1) and flexion-extension (Video2) motions that are used to train subjects to follow during the scan:}\\
Video1: link will be provided by Publisher \\
Video2: link will be provided by Publisher
\newline

\begin{figure}[h]
     \centering
      \caption{
      {\bf Unconstrained wrist positioning in the flex coil during scanning of static and moving images.}}
        \includegraphics[width=17cm]{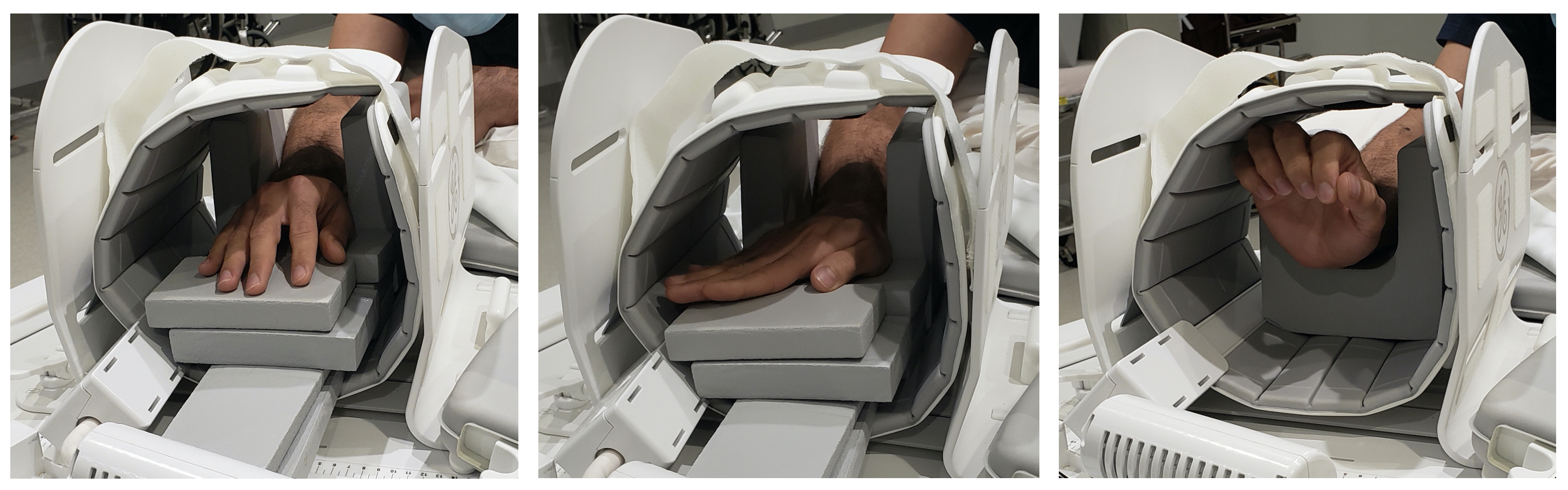}
        \label{wrist}
\end{figure}
\begin{figure}[h!]
\centering
 \caption{
 {\bf Orthogonal planes of (a) 3D-SPGR, (b) ZTE, and (c) LAVA Flex MRI modalities of wrist carpal bones of subject 1. Red arrow points the abnormality observed in capitate.}}
 \includegraphics[width=18cm]{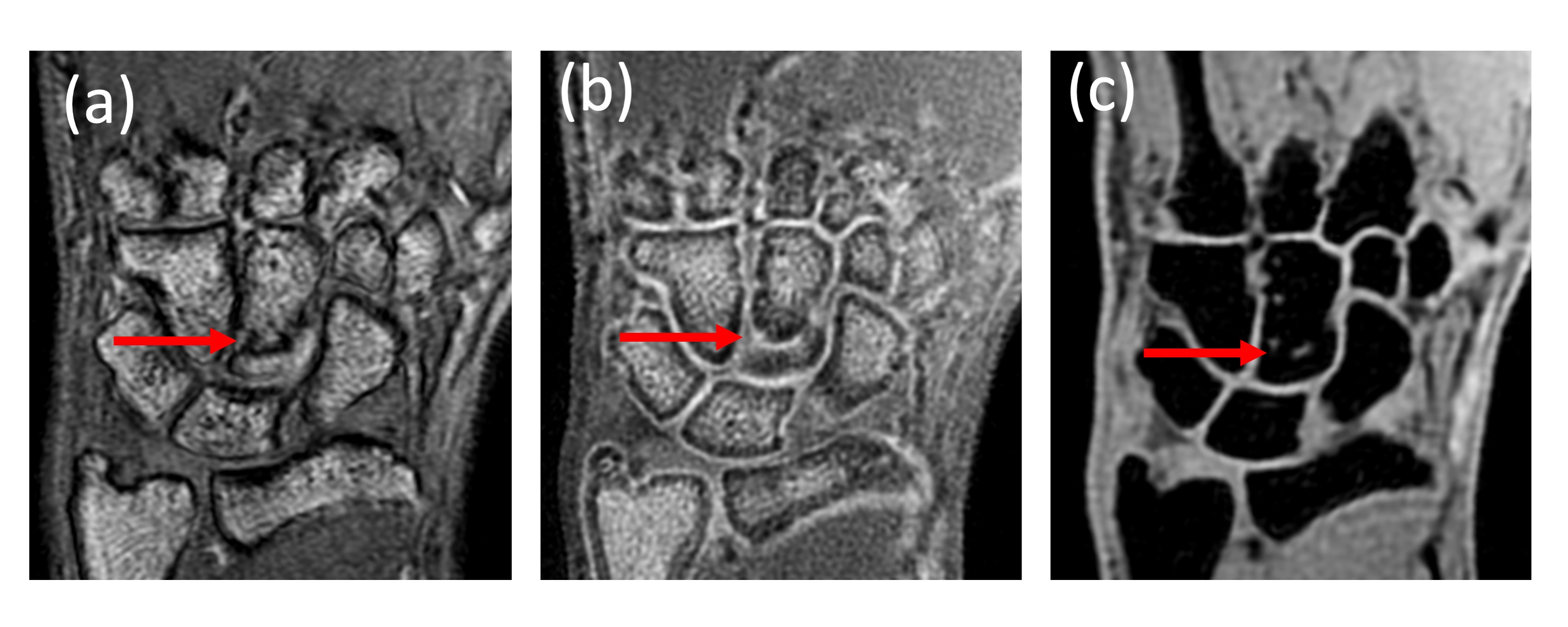}
\label{abnormality}
\end{figure}

\begin{figure}[]
     \centering
     \caption{
     {\bf Kinematic profiles of center-of-mass displacements of simultaneously registered  scaphoid, capitate, and lunate (blue) and the combined canter-of-mass of individually registered bones (red).}}
        \includegraphics[width=18cm]{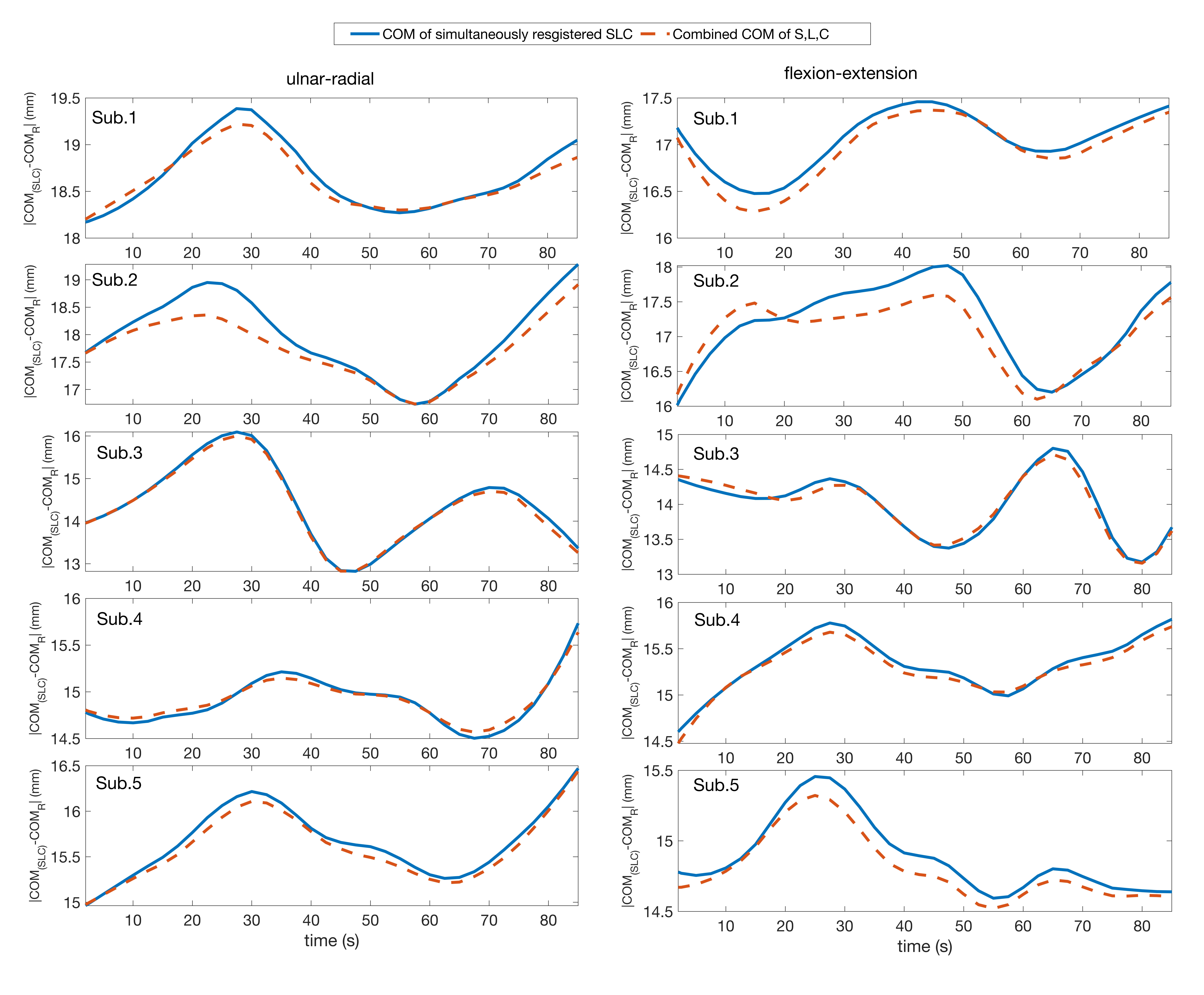}
        \label{combinedCOM}
\end{figure}


\begin{thebibliography}{10}

\bibitem{Muhle1999}
Muhle C, Brossmann J, Heller M.
\newblock Kinematic CT and MR imaging of the patellofemoral joint.
\newblock European Radiology 1999;\hspace{0pt}93:508--518.

\bibitem{Johnson2013}
Johnson JE, Lee P, McIff TE, Toby EB, Fischer KJ.
\newblock Scapholunate ligament injury adversely alters in vivo wrist joint
  mechanics: An MRI-based modeling study.
\newblock Journal of Orthopaedic Research 2013;\hspace{0pt}319:1455--1460.

\bibitem{Li2004}
Li G, Wuerz TH, DeFrate LE.
\newblock {Feasibility of Using Orthogonal Fluoroscopic Images to Measure In
  Vivo Joint Kinematics }.
\newblock Journal of Biomechanical Engineering 2004;\hspace{0pt}1262:313--318.

\bibitem{Teixeira2015}
{Gondim Teixeira} PA, Gervaise A, Louis M, Lecocq S, Raymond A, Aptel S, Blum
  A.
\newblock Musculoskeletal wide detector CT: Principles, techniques and
  applications in clinical practice and research.
\newblock European Journal of Radiology 2015;\hspace{0pt}845:892--900.

\bibitem{Langner2015}
Langner I, Fischer S, Eisenschenk A, Langner S.
\newblock Cine MRI: a new approach to the diagnosis of scapholunate
  dissociation.
\newblock Skeletal Radiology 2015;\hspace{0pt}448:1103--1110.

\bibitem{Stromps2018}
Stromps JP, Eschweiler J, Knobe M, Rennekampff HO, Radermacher K, Pallua N.
\newblock Impact of scapholunate dissociation on human wrist kinematics.
\newblock Journal of Hand Surgery (European Volume)
  2018;\hspace{0pt}432:179--186.
\newblock PMID: 26307143.

\bibitem{You2001}
You BM, Siy P, Anderst W, Tashman S.
\newblock In vivo measurement of 3-D skeletal kinematics from sequences of
  biplane radiographs: Application to knee kinematics.
\newblock IEEE Transactions on Medical Imaging 2001;\hspace{0pt}206:514--525.

\bibitem{SENNWALD1993805}
Sennwald GR, Zdravkovic V, Kern HP, Jacob HA.
\newblock Kinematics of the wrist and its ligaments.
\newblock The Journal of Hand Surgery 1993;\hspace{0pt}185:805--814.

\bibitem{Bateni2013}
Bateni CP, Bartolotta RJ, Richardson ML, Mulcahy H, Allan CH.
\newblock Imaging Key Wrist Ligaments: What the Surgeon Needs the Radiologist
  to Know.
\newblock American Journal of Roentgenology 2013;\hspace{0pt}2005:1089--1095.

\bibitem{Feipel1999}
Feipel V, Rooze M.
\newblock Three-dimensional motion patterns of the carpal bones: an in vivo
  study using three-dimensional computed tomography and clinical applications.
\newblock Surgical and Radiologic Anatomy 1999;\hspace{0pt}212:125--131.

\bibitem{Sikora2019}
Sikora SK, Tham SK, Harvey JN, Garcia-Elias M, Goldring T, Rotstein AH, Ek ET.
\newblock The Twist X-Ray: A Novel Test for Dynamic Scapholunate Instability.
\newblock J Wrist Surg 01.02.2019;\hspace{0pt}0801:061--065.
\newblock 061.

\bibitem{Rainbow2013}
Rainbow MJ, Kamal RN, Leventhal E, Akelman E, Moore DC, Wolfe SW, Crisco JJ.
\newblock In vivo kinematics of the scaphoid, lunate, capitate, and third
  metacarpal in extreme wrist flexion and extension.
\newblock The Journal of hand surgery 2013;\hspace{0pt}382:278--288.
\newblock 23266007[pmid].

\bibitem{Best2019}
Best GM, Mack ZE, Pichora DR, Crisco JJ, Kamal RN, Rainbow MJ.
\newblock Differences in the Rotation Axes of the Scapholunate Joint During
  Flexion-Extension and Radial-Ulnar Deviation Motions.
\newblock The Journal of hand surgery 2019;\hspace{0pt}449:772--778.
\newblock 31300230[pmid].

\bibitem{Roo2019}
de~Roo MGA, Muurling M, Dobbe JGG, Brinkhorst ME, Streekstra GJ, Strackee SD.
\newblock A four-dimensional-CT study of in vivo scapholunate rotation axes:
  possible implications for scapholunate ligament reconstruction.
\newblock Journal of Hand Surgery (European Volume)
  2019;\hspace{0pt}445:479--487.
\newblock PMID: 30813846.

\bibitem{Zhao2015}
Zhao K, Breighner R, Holmes I David, Leng S, McCollough C, An KN.
\newblock {A Technique for Quantifying Wrist Motion Using Four-Dimensional
  Computed Tomography: Approach and Validation}.
\newblock Journal of Biomechanical Engineering 2015;\hspace{0pt}1377.
\newblock 074501.

\bibitem{Beek2004}
Beek M, Small CF, Csongvay S, Sellens RW, Ellis RE, Pichora DR.
\newblock Wrist Kinematics from Computed Tomography Data.
\newblock Medical Image Computing and Computer-Assisted Intervention -- MICCAI
  2004.
\newblock 2004 1040--1041.

\bibitem{Crisco2005}
Crisco JJ, Coburn JC, Moore DC, Akelman E, Weiss APC, Wolfe SW.
\newblock In Vivo Radiocarpal Kinematics and the Dart Thrower's Motion.
\newblock JBJS 2005;\hspace{0pt}8712.

\bibitem{FOSTER2019}
Foster BH, Shaw CB, Boutin RD, Joshi AA, Bayne CO, Szabo RM, Chaudhari AJ.
\newblock A principal component analysis-based framework for statistical
  modeling of bone displacement during wrist maneuvers.
\newblock Journal of Biomechanics 2019;\hspace{0pt}85:173--181.

\bibitem{Henrichon2020}
Henrichon SS, Foster BH, Shaw C, Bayne CO, Szabo RM, Chaudhari AJ, Boutin RD.
\newblock Dynamic MRI of the wrist in less than 20 seconds: normal midcarpal
  motion and reader reliability.
\newblock Skeletal Radiology 2020;\hspace{0pt}492:241--248.

\bibitem{LAVAFlex}
Li XH, Zhu J, Zhang XM, Ji YF, Chen TW, Huang XH, Yang L, Zeng NL.
\newblock Abdominal MRI at 3.0 T: LAVA-flex compared with conventional fat
  suppression T1-weighted images.
\newblock Journal of Magnetic Resonance Imaging 2014;\hspace{0pt}401:58--66.

\bibitem{FERRANTE2017}
Ferrante E, Paragios N.
\newblock Slice-to-volume medical image registration: A survey.
\newblock Medical Image Analysis 2017;\hspace{0pt}39:101--123.

\bibitem{OHIF}
Urban T, Ziegler E, Lewis R, Hafey C, Sadow C, Van~den Abbeele AD, Harris GJ.
\newblock LesionTracker: Extensible Open-Source Zero-Footprint Web Viewer for
  Cancer Imaging Research and Clinical Trials.
\newblock Cancer Research 2017;\hspace{0pt}7721:e119--e122.

\bibitem{matlab}
MATLAB.
\newblock version 9.8.0.1396136 (R2020a).
\newblock Natick, Massachusetts: The MathWorks Inc., 2020.

\end{thebibliography}
\end{document}